


\font\bigbold=cmbx12
\font\ninerm=cmr9
\font\eightrm=cmr8
\font\sixrm=cmr6
\font\fiverm=cmr5
\font\ninebf=cmbx9
\font\eightbf=cmbx8
\font\sixbf=cmbx6
\font\fivebf=cmbx5
\font\ninei=cmmi9  \skewchar\ninei='177
\font\eighti=cmmi8  \skewchar\eighti='177
\font\sixi=cmmi6    \skewchar\sixi='177
\font\fivei=cmmi5
\font\ninesy=cmsy9 \skewchar\ninesy='60
\font\eightsy=cmsy8 \skewchar\eightsy='60
\font\sixsy=cmsy6   \skewchar\sixsy='60
\font\fivesy=cmsy5
\font\nineit=cmti9
\font\eightit=cmti8
\font\ninesl=cmsl9
\font\eightsl=cmsl8
\font\ninett=cmtt9
\font\eighttt=cmtt8
\font\tenfrak=eufm10
\font\ninefrak=eufm9
\font\eightfrak=eufm8
\font\sevenfrak=eufm7
\font\fivefrak=eufm5
\font\tenbb=msbm10
\font\ninebb=msbm9
\font\eightbb=msbm8
\font\sevenbb=msbm7
\font\fivebb=msbm5
\font\tensmc=cmcsc10


\newfam\bbfam
\textfont\bbfam=\tenbb
\scriptfont\bbfam=\sevenbb
\scriptscriptfont\bbfam=\fivebb
\def\Bbb{\fam\bbfam}

\newfam\frakfam
\textfont\frakfam=\tenfrak
\scriptfont\frakfam=\sevenfrak
\scriptscriptfont\frakfam=\fivefrak
\def\frak{\fam\frakfam}

\def\smc{\tensmc}


\def\eightpoint{%
\textfont0=\eightrm   \scriptfont0=\sixrm
\scriptscriptfont0=\fiverm  \def\rm{\fam0\eightrm}%
\textfont1=\eighti   \scriptfont1=\sixi
\scriptscriptfont1=\fivei  \def\oldstyle{\fam1\eighti}%
\textfont2=\eightsy   \scriptfont2=\sixsy
\scriptscriptfont2=\fivesy
\textfont\itfam=\eightit  \def\it{\fam\itfam\eightit}%
\textfont\slfam=\eightsl  \def\sl{\fam\slfam\eightsl}%
\textfont\ttfam=\eighttt  \def\tt{\fam\ttfam\eighttt}%
\textfont\frakfam=\eightfrak \def\frak{\fam\frakfam\eightfrak}%
\textfont\bbfam=\eightbb  \def\Bbb{\fam\bbfam\eightbb}%
\textfont\bffam=\eightbf   \scriptfont\bffam=\sixbf
\scriptscriptfont\bffam=\fivebf  \def\bf{\fam\bffam\eightbf}%
\abovedisplayskip=9pt plus 2pt minus 6pt
\belowdisplayskip=\abovedisplayskip
\abovedisplayshortskip=0pt plus 2pt
\belowdisplayshortskip=5pt plus2pt minus 3pt
\smallskipamount=2pt plus 1pt minus 1pt
\medskipamount=4pt plus 2pt minus 2pt
\bigskipamount=9pt plus4pt minus 4pt
\setbox\strutbox=\hbox{\vrule height 7pt depth 2pt width 0pt}%
\normalbaselineskip=9pt \normalbaselines
\rm}


\def\ninepoint{%
\textfont0=\ninerm   \scriptfont0=\sixrm
\scriptscriptfont0=\fiverm  \def\rm{\fam0\ninerm}%
\textfont1=\ninei   \scriptfont1=\sixi
\scriptscriptfont1=\fivei  \def\oldstyle{\fam1\ninei}%
\textfont2=\ninesy   \scriptfont2=\sixsy
\scriptscriptfont2=\fivesy
\textfont\itfam=\nineit  \def\it{\fam\itfam\nineit}%
\textfont\slfam=\ninesl  \def\sl{\fam\slfam\ninesl}%
\textfont\ttfam=\ninett  \def\tt{\fam\ttfam\ninett}%
\textfont\frakfam=\ninefrak \def\frak{\fam\frakfam\ninefrak}%
\textfont\bbfam=\ninebb  \def\Bbb{\fam\bbfam\ninebb}%
\textfont\bffam=\ninebf   \scriptfont\bffam=\sixbf
\scriptscriptfont\bffam=\fivebf  \def\bf{\fam\bffam\ninebf}%
\abovedisplayskip=10pt plus 2pt minus 6pt
\belowdisplayskip=\abovedisplayskip
\abovedisplayshortskip=0pt plus 2pt
\belowdisplayshortskip=5pt plus2pt minus 3pt
\smallskipamount=2pt plus 1pt minus 1pt
\medskipamount=4pt plus 2pt minus 2pt
\bigskipamount=10pt plus4pt minus 4pt
\setbox\strutbox=\hbox{\vrule height 7pt depth 2pt width 0pt}%
\normalbaselineskip=10pt \normalbaselines
\rm}


\def\pagewidth#1{\hsize= #1}
\def\pageheight#1{\vsize= #1}
\def\hcorrection#1{\advance\hoffset by #1}
\def\vcorrection#1{\advance\voffset by #1}

\newif\iftitlepage   \titlepagetrue               
\newtoks\titlepagefoot     \titlepagefoot={\hfil} 
\newtoks\otherpagesfoot    \otherpagesfoot={\hfil\tenrm\folio\hfil}
\footline={\iftitlepage\the\titlepagefoot\global\titlepagefalse
           \else\the\otherpagesfoot\fi}

\font\extra=cmss10 scaled \magstep0
\setbox1 = \hbox{{{\extra R}}}
\setbox2 = \hbox{{{\extra I}}}
\setbox3 = \hbox{{{\extra C}}}
\setbox4 = \hbox{{{\extra Z}}}
\setbox5 = \hbox{{{\extra N}}}

\def\RRR{{{\extra R}}\hskip-\wd1\hskip2.0 true pt{{\extra I}}\hskip-\wd2
\hskip-2.0 true pt\hskip\wd1}
\def\Real{\hbox{{\extra\RRR}}}    

\def\C{{{\extra C}}\hskip-\wd3\hskip2.5 true pt{{\extra I}}\hskip-\wd2
\hskip-2.5 true pt\hskip\wd3}
\def\Complex{\hbox{{\extra\C}}}   

\def\Z{{{\extra Z}}\hskip-\wd4\hskip 2.5 true pt{{\extra Z}}}
\def\Zed{\hbox{{\extra\Z}}}       

\def\N{{{\extra N}}\hskip-\wd5\hskip -1.5 true pt{{\extra I}}}



\def\H{{\cal H}}
\def\M{M}
\def\RM{{M}^{\rm red}}
\def\G{{\cal G}}

\def\a{\alpha}
\def\b{\beta}

\def\tr{\hbox{tr}}


\def\abstract#1{{\parindent=30pt\narrower\noindent\ninepoint\openup
2pt #1\par}}


\newcount\notenumber\notenumber=1
\def\note#1
{\unskip\footnote{$^{\the\notenumber}$}
{\eightpoint\openup 1pt #1}
\global\advance\notenumber by 1}


\def\frac#1#2{{#1\over#2}}

\def\({\left(}
\def\){\right)}
\def\<{\langle}
\def\>{\rangle}

\def\pmb#1{\setbox0=\hbox{$#1$}%
   \kern-.025em\copy0\kern-\wd0
   \kern.05em\copy0\kern-\wd0
   \kern-.025em\raise.0433em\box0 }


\global\newcount\secno \global\secno=0
\global\newcount\meqno \global\meqno=1
\global\newcount\appno \global\appno=0
\newwrite\eqmac
\def\romappno{\ifcase\appno\or A\or B\or C\or D\or E\or F\or G\or H
\or I\or J\or K\or L\or M\or N\or O\or P\or Q\or R\or S\or T\or U\or
V\or W\or X\or Y\or Z\fi}
\def\eqn#1{
        \ifnum\secno>0
            \eqno(\the\secno.\the\meqno)\xdef#1{\the\secno.\the\meqno}
          \else\ifnum\appno>0
            \eqno({\rm\romappno}.\the\meqno)\xdef#1{{\rm\romappno}.\the\meqno}
          \else
            \eqno(\the\meqno)\xdef#1{\the\meqno}
          \fi
        \fi
\global\advance\meqno by1 }


\global\newcount\refno
\global\refno=1 \newwrite\reffile
\newwrite\refmac
\newlinechar=`\^^J
\def\ref#1#2{\the\refno\nref#1{#2}}
\def\nref#1#2{\xdef#1{\the\refno}
\ifnum\refno=1\immediate\openout\reffile=refs.tmp\fi
\immediate\write\reffile{
     \noexpand\item{[\noexpand#1]\ }#2\noexpand\nobreak}
     \immediate\write\refmac{\def\noexpand#1{\the\refno}}
   \global\advance\refno by1}
\def\semi{;\hfil\noexpand\break ^^J}
\def\nl{\hfil\noexpand\break ^^J}
\def\refn#1#2{\nref#1{#2}}

\def\pl#1#2#3{{ Phys. Lett.} {\bf {#1}B} (19{#2}) #3}

\def\prD#1#2#3{{ Phys. Rev.} {\bf D{#1}} (19{#2}) #3}

\def\prp#1#2#3{{ Phys. Rep.} {\bf {#1}} (19{#2}) #3}

%
%

{
\ninepoint
\refn\DJ
{
E.\ D'Hoker, R.\ Jackiw,
{\it Classical and quantal Liouville field theory},
\prD{26}{82}{3517-3542}.
}
\refn\FORTW
{
See, for example,
L.\ Feh\'er, L.\ O'Raifeartaigh, P.\ Ruelle, I.\ Tsutsui, A.\ Wipf,
{\it
On Hamiltonian reductions of the Wess-Zumino-Novikov-Witten theories},
\prp{222}{92}{1-64}.
}
\refn\FT
{
L.\ Feh\'er, I.\ Tsutsui,
{\it  Regularization of  Toda lattices by Hamiltonian reduction},
 preprint, INS-Rep.-1123 (hep-th/9511118).
}                              
\refn\TF
{
I.\ Tsutsui, L.\ Feh\'er,
{\it Global aspects of the WZNW reduction to Toda theories},
Prog.\ Theor.\ Phys.\ Suppl.\ {\bf 118} (1995) 173-190.
} 
\refn\RSTS
{
A.G.\ Reyman, M.A.\ Semenov-Tian-Shansky,
{\it Reduction of Hamiltonian systems, Affine Lie Algebras and 
Lax Equations}, Invent. Math. {\bf 54} (1979) 81-100.
}
\refn\F
{
T.\ F\"ul\"op,
{\it Reduced $SL(2,\Real)$ WZNW quantum mechanics},
ITP Budapest Report 509, hep-th/9502145 (to appear in J.\ Math.\ Phys.)
}.
\refn\He
{
S.\  Helgason,  {\sl Differential Geometry,
Lie Groups and Symmetric Spaces}, Chapter IX,
Academic Press, New York, 1978.
}
\refn\DS
{
V.G.\ Drinfeld, V.V.\ Sokolov,
{\it Lie algebras and equations of KdV type},
J.\ Sov.\ Math.\ {\bf 30} (1985) 1975-2036.
}
\refn\RS
{
See, for example,
M.\ Reed, B.\ Simon, {\sl Methods of Modern Mathematical
Physics I: Functional Analysis\/},
Academic Press, New York, 1972.
}
\refn\ORT
{
L.\ O'Raifeartaigh, P.\ Ruelle, I.\ Tsutsui,
{\it Quantum Equivalence
 of Constrained WZNW and Toda Theories},
\pl{258}{91}{359-363}.
}                                           
\refn\N
{
M.A.\ Naimark, {\sl Linear Differential Operators: Part II\/},
Frederick Unger, New York, 1968.
}
\refn\W
{
G.N.\ Watson, {\sl Theory of Bessel Functions\/},
Springer-Verlag, Singapore, 1989.
}
}


\pageheight{23cm}
\pagewidth{15.7cm}
\hcorrection{-1mm}
\magnification= \magstep1
\baselineskip=14.6pt plus 1pt minus 1pt
\parskip=5pt plus 1pt minus 1pt
\tolerance 8000


%
%


\null
{\baselineskip=10pt
\rightline{INS-Rep.-1124\break}
\rightline{January 1996\break}
\vfill
}
\centerline{\bigbold 
Quantum mechanical Liouville model with attractive potential}
\vskip 30pt
\centerline{\smc Hiroyuki Kobayashi\note
{e-mail: kobayasi@ins.u-tokyo.ac.jp}
}
\vskip 5pt
\centerline
{\smc and}
\vskip 5pt
\centerline{\smc Izumi Tsutsui\note
{e-mail: tsutsui@ins.u-tokyo.ac.jp}
}

\vskip 8pt
{\baselineskip=12pt
\centerline{\it Institute for Nuclear Study}
\centerline{\it University of Tokyo}
\centerline{\it Midori-cho, Tanashi-shi, Tokyo 188}
\centerline{\it Japan}
}
\vskip 30pt
\abstract{%
{\bf Abstract.}\quad 
We study the quantum mechanical Liouville model with attractive potential
which is obtained by Hamiltonian symmetry reduction from the system
of a free particle on $SL(2, \Real)$.  The classical reduced system
consists of a pair of Liouville
subsystems which are `glued together' in such a way that the singularity
of the Hamiltonian flow is regularized.  It is shown that the quantum
theory of this reduced system
is labelled by an angle parameter $\theta \in [\,0,2\pi)$
characterizing the self-adjoint extensions of the Hamiltonian and hence
the energy spectrum.
There exists a probability flow between the two Liouville subsystems,
demonstrating that the two subsystems are also 
`connected' 
quantum mechanically, even though 
all the wave functions in the 
Hilbert space vanish at the junction.
}

\vfill\eject


\noindent \secno=1 \meqno=1

\centerline
{\bf 1. Introduction}
\medskip
\noindent
The Liouville model
has been under intense scrutiny in recent years,
due to its relation to two dimensional quantum gravity, which 
is important in
the theory of the world sheet of string theories.  The model,
which is solvable classically, has proved to be a rich source  
for developing techniques as well as 
for probing the universal features of
quantum gravity in higher dimensions.  However, the problem is that
despite many fruitful achievements the model still resists a full  
understanding as a quantum theory.  It is therefore heartening
to observe
that its toy model version obtained
by ignoring the space dimension governed by the Hamiltonian, 
$$
H(\pi, x) = \pi^2 + \mu e^{2x},
\eqn\lvham
$$
can be solved completely for $\mu > 0$ even quantum mechanically.
The quantum mechanical Liouville model possesses a continuous
energy spectrum and its eigenstates are 
given by modified Bessel functions [\DJ].  
A somewhat peculiar
aspect of the quantum theory is that it has no vacuum state even 
though the energy is bounded from below, an aspect that
stems from the repulsive exponential potential, which 
has no minimum.
Now one might ask what happens 
if we replace the potential `wall' with a `well' by putting $\mu < 0$.
This attractive exponential potential
will undoubtedly give rise to difficulty in quantization
because the energy would then be unbounded from below without any
bound states, indicating the quantum
instability of the system.  This is a reflection
of the classical instability whereby the particle
sinks indefinitely fast into the well, as
the classical solution develops a singularity 
and blows up at some finite time if $\mu < 0$.   
This is perhaps the reason why the quantum theory of 
the Liouville model
with attractive potential has not been considered seriously so far.

Meanwhile, we have found in the study of ${\cal W}$-algebras 
and the generalized KdV systems 
that the Toda field theory, obtained by
Hamiltonian symmetry reduction [\FORTW] 
from the Wess-Zumino-Novikov-Witten
model based on a Lie group $G$, 
has a certain global structure [\FT].  More precisely,
the reduced theory is not merely a Toda theory but
consists of a multiple of Toda theories as subsystems having
both repulsive and attractive potentials in general.  In particular,
for $G = SL(2, \Real)$ we have a pair of Liouville models
as subsystems in the reduced system, both of which have an either
repulsive or attractive potential depending on the reduction 
performed.    
The interesting observation made there [\FT, \TF] (see also [\RSTS])
was that,
in the toy model version where the Toda field theories
become the Toda lattices,
the singular classical solutions 
that arise in the Toda lattices are
regularized automatically by the Hamiltonian reduction.
An intuitive picture of 
the regularization may be gained by considering the simple 
case $G = SL(2, \Real)$ where we get two Liouville (toy) models   
which are `glued together' by identifying the limits
$x \rightarrow \infty$ of the two models (see Fig. 1).
The singular solution in one Liouville model is regularized 
by continuing it in time 
to the solution in the other Liouville model, causing
the particle to oscillate between the two subsystems.
This observation motivated us to investigate more closely
the Liouville model with attractive potential,
now regularized in the above sense, to see if any sensible quantization
is possible.  

A first step in this direction was made 
in a paper by F\"ul\"op [\F], where, 
like the former repulsive potential case, the theory
is solved completely at the quantum level yielding
Bessel functions as energy eigenstates.  The salient
result of [\F] is that the spectrum is discrete,
which is perhaps a natural consequence of the regularized
classical solutions being oscillations,
and that there are {\it inequivalent quantizations} 
characterized by certain parameters
specifying the self-adjoint quantum
Hamiltonian and hence the spectrum.  
However, the argument
in [\F] appears to be 
unnecessarily complicated 
at a few crucial points, especially when the
self-adjoint Hamiltonians are constructed
over the entire reduced system.  The aim of this paper is to
present a quantization approach which is much simpler and more direct
in these points, and to furnish 
a complete version of 
the (regularized) quantum mechanical Liouville model with attractive
potential.  
We shall find that, as in [\F], there arise  
inequivalent quantizations but they can be characterized just by 
an angle parameter $\theta \in [\, 0, 2\pi)$, and that the
discrete energy spectrum obtained  
turns out to be different from that of [\F].  
We also see more naturally 
a probability
flow between the two Liouville subsystems, a fact 
demonstrating that these
subsystems are also connected quantum mechanically.

The plan of the paper is as follows: To make the paper self-contained,
in section 2 we provide a necessary
background for the classical reduced system.  
Then in section 3 we present a quantum 
theory of the reduced system, which
is a combined system of two Liouville models with attractive potential.  
The final section is devoted to discussion.  We provide
two appendices; Appendix A for a brief review of the general theory
of self-adjoint extensions of symmetric operators, and Appendix B
for a collection of formulae involving Bessel functions 
used in the text.


\noindent \secno=2 \meqno=1

\bigskip
\medskip
\centerline
{\bf 2. Hamiltonian reduction and the global structure}
\medskip
\noindent
In order to set the scene for the system for 
which we discuss the quantization,
we here recall the
Hamiltonian symmetry reduction [\FORTW] 
which leads to
the system of regularized Liouville models, together with the global
description of the system developed recently [\FT].

The reduction is the special case $n = 2$ 
of the Hamiltonian
reduction that yields a multiple ($2^{n-1}$) of  
open, finite Toda lattices from 
the free particle system on the group $G=SL(n,\Real )$.  
In the reduced system these Toda lattices, which have in general
both repulsive and attractive potentials, are `glued together'
in such a way that no singularity arises.
In this sense the reduction provides 
a natural means to regularize the singularities
which exist in those Toda lattices that have 
attractive potentials.  (This regularization is an example of
a more general idea put forwarded originally in [\RSTS].)

\medskip
\noindent{\bf 2.1. Classical Hamiltonian reduction}

\noindent
The free particle on a semisimple Lie group $G$ 
is described by the Hamiltonian system $(M,\Omega,H)$ 
in the following way.
The phase space $\M$ is the cotangent bundle of the group,
$$
M = T^*G \simeq G \times \G = \left\{(g,J)\, 
\vert\,\, g \in G, \, J \in {\cal G}\,\right\},
\eqn\fps
$$
where $\G$ is the Lie algebra of $G$ (in our case $\G = sl(2,\Real)$),
which is identified
with its dual $\G^*$ using the scalar product.
The fundamental Poisson brackets are
$$
\{g_{ij}\, , g_{kl}\} = 0, \quad
      \{g_{ij}\, , \tr(T^aJ) \} = (T^a g)_{ij}, \quad
\{\tr(T^aJ)\, , \tr (T^bJ)\} = \tr ([T^a, T^b] J),
\eqn\fpb
$$
where $\{T^a\}$ is a basis of $\G$.  For $\G = sl(2,\Real)$
we may take 
$$
T^+ = \pmatrix{
0 & 1 \cr
0 & 0 \cr}, \qquad
T^0 = \pmatrix{
1 & 0 \cr
0 & -1 \cr}, \qquad
T^- = \pmatrix{
0 & 0 \cr
1 & 0 \cr}. 
\eqn\basis
$$
The Poisson brackets (\fpb) derive from the symplectic form,
$$
\Omega = d\, \tr \left( J dg\, g^{-1}\right).
\eqn\sf
$$
The Hamiltonian is given by 
$$
H = {1\over2} \tr\left( J^2\right),
\eqn\ham
$$
which leads to the dynamics
$${d g\over dt} = \{ g \, , H\} = Jg, \qquad 
{d J\over dt} = \{ J \, , H\} = 0,
\eqn\heq
$$
and hence yields the geodesic equation 
${d\over{dt}}({{dg}\over{dt}} \, g^{-1}) = 0$
on the group manifold $G$.
We note that 
$J$ is the infinitesimal generator 
for the action of $G$ on $\M$ defined by left translations, while the
action of $G$ defined by right translations is generated by
$\tilde J: \M\rightarrow \G$ where
$$
\tilde J(g,J) := -g^{-1} Jg.
\eqn\rcur
$$

Let us now decompose the Lie algebra $\G = sl(2, \Real)$ 
into the subalgebras of
strictly upper triangular, diagonal, 
and strictly lower triangular traceless 
matrices, that is, those subalgebras spanned by the basis (\basis),
$$
\G = \G_{+} + \G_0 +\G_{-}.
\eqn\decalg
$$
Then we consider symmetry reduction based on the subgroup
$N := N_+\times N_- \subset G$, where 
$N_\pm = \hbox{exp}(\G_\pm)$,
which acts on the phase space $\M$ according to
$$
(n_+,n_-): (g,J)\mapsto (n_+ g n_-^{-1}, n_+Jn_+^{-1}),
\qquad \forall (n_+,n_-)\in N, \quad (g,J)\in \M.
\eqn\naction
$$
The symmetry reduction is performed by decomposing
the generators, $J = J_+ + J_0 + J_-$ and 
$\tilde J= \tilde J_+ + \tilde J_0 + \tilde J_-$, 
according to (\decalg) and 
then fixing the value of the momentum map 
$\Phi(g,J):= ( J_{-}, \tilde J_{+})$ as
$$
\Phi(g,J)
= \left( I_{-}, - I_{+}\right),
\eqn\momentmap
$$
where $I_- := \nu^- T_-$ and $I_+ := \nu^+ T_+$ 
are nonvanishing constant matrices ($\nu^\pm \ne 0$) belonging to
$\G_-$ and $\G_+$, respectively.
The reduced phase space is obtained as the factor space
$$
\M^{\rm red}(I_-,I_+)=\M^{\rm c}(I_-,I_+)/N,
\qquad\hbox{where}\quad 
\M^{\rm c}(I_-,I_+):=\Phi^{-1}\left(I_-, -I_+\right).
\eqn\redps
$$
In Dirac's terminology, this Hamiltonian reduction amounts to
imposing the first class constraints, 
$J_{-}=I_-$ and $\tilde J_{+}=-I_+$, 
defining $\M^{\rm c}\subset \M$, 
and getting the reduced phase space by fixing
the gauge associated with the symmetry group $N$
generated by the constraints; hence (\redps).

Now the {\it Bruhat (Gelfand-Naimark) decomposition}
for semisimple
Lie groups [\He] allows us to write $G = SL(2, \Real)$ as
$$
G= G_e \cup G_{-e} \cup G_{\rm low}\qquad \hbox{(disjoint union)},
\eqn\bruhat
$$
where 
$$
G_{\pm e} := \pm N_+ A N_- \quad \hbox{with} 
\quad  A:=\exp\left(\G_0\right).
\eqn\submfd
$$
The two `cells', $G_e$ and $G_{-e}$, are open submanifolds in $G$
containing $e$ and $-e$ ($e \in G$ is the identity element), respectively
(and their union is dense in $G$), while $G_{\rm low}$ is 
the union of `borders', i.e., certain lower dimensional
submanifolds of $G$.  Correspondingly, any element $g \in 
G_e \cup G_{-e} \subset SL(2, \Real)$
admits the unique decomposition in the form,
$$
g = \pm n_+ e^q n_- = \pm   
\pmatrix{
1 & a \cr
0 & 1 \cr}
\pmatrix{
e^x & 0 \cr
0 & e^{-x} \cr}
\pmatrix{
1 & 0 \cr
c & 1 \cr},
\eqn\decomp
$$
where we put $q = x T_0$ and $a$, $c$, $x \in \Real$.
The two cells $G_{\pm e}$ are
in fact the open submanifolds of determinant one
matrices with $\hbox{sign}(g_{22}) = \pm $, whereas
$G_{\rm low}$
consists of those matrices with $g_{22} = 0$.

The Bruhat decomposition naturally induces 
the decomposition of the phase space $\M=T^*G$ as
$\M = \M_e \cup M_{-e} \cup \M_{\rm low}$.
Since this decomposition is invariant 
under the action of the symmetry group $N$, we have the corresponding
decomposition of $M^{\rm c} = M^{\rm c}_e \cup \M^{\rm c}_{-e}
\cup \M^{\rm c}_{\rm low}$ as well, 
which in turn induces the decomposition
of the reduced phase space,
$$
\M^{\rm red} = M^{\rm red}_e \cup \M^{\rm red}_{-e} 
    \cup \M^{\rm red}_{\rm low}
\qquad\hbox{(disjoint union)}.
\eqn\bruhatrps
$$
We note that 
$\M^{\rm red}_{\pm e}$ are open submanifolds in $\M^{\rm red}$
and $\M^{\rm red}_{\rm low}$ is
a union of lower dimensional submanifolds.  In other words, 
the Bruhat decomposition introduces 
the cell-structure in the reduced phase space.
We now show that each of the subsystems associated with the two cells  
$\M^{\rm red}_{\pm e}$ is a Liouville model. 
Indeed, since the submanifolds $\M^{\rm c}_{\pm e}$ are
$$
\M^{\rm c}_{\pm e} = \left\{\,(g ,J)\,\vert\, g = \pm n_+ e^q n_-,\,\,
 n_\pm \in N_\pm,\,\, q\in \G_0,
\,\,\, J_{-}=I_-,\,\, \left(g^{-1} J g\right)_{+}=I_+\,\right\},
\eqn\submc
$$
we see that
$\M_{\pm e}^{\rm red}=\M^{\rm c}_{\pm e}/(N_+\times N_-)$ 
are given by the local gauge section,
$$
\M^{\rm red}_{\pm e} = \left\{\,(\pm e^q ,J)\,\vert\,   q\in \G_0,
\quad J_{-} = 
I_-,\,\, \left(e^{-q} J e^{q}\right)_{+}=I_+\,\right\}.
\eqn\lgs
$$
The condition in (\lgs) can easily be solved for $J$:
$$
J = J_{\pm e}(q,p) := I_- + p + e^q I_+ e^{-q},
\eqn\soluj
$$
where $q$, $p \in \G_0$.  Thus we may write
$$
\M^{\rm red}_{\pm e}  
= \left\{\,(\pm e^q ,J_{\pm e}(q,p) )\, \vert\,  (q,p) 
\in \G_0 \times \G_0\, \right\},
\eqn\redpssvd
$$
or simply $\M^{\rm red}_{\pm e} \simeq \G_0 \times \G_0 = \Real^2$.
By evaluating the symplectic form 
(\sf) on the reduced phase space (\redpssvd), we find 
$$
\Omega^{\rm red}_{\pm e} =  d\, 
\tr \left( J_{\pm e}\, d(e^q)\, e^{-q} \right) 
= d\, \tr\left(p dq\right)  = 2 d\pi dx,
\eqn\redsf
$$
where we put $p = \pi T_0$ with $\pi \in \Real$.
Similarly, from (\ham) the reduced Hamiltonian 
turns out to be
$$
H_{\pm e}^{\rm red}(J_{\pm e}) = {1\over 2}\tr\left(J_{\pm e}^2\right)
= {1\over 2}\tr(p^2) + \mu \, \tr (T_- e^q T_+ e^{-q})
= \pi^2 + \mu e^{2x},
\eqn\redham
$$
where $\mu = \nu^+\nu^-$.  Since the triples
$(\RM_{\pm e}, \Omega^{\rm red}_{\pm e}, H_{\pm e}^{\rm red})$
are no other than that of the Liouvile model, 
we conclude that the reduced
system obtained by the Hamiltonian reduction 
possesses two, identical Liouville models as subsystems.

\medskip
\noindent
{\bf 2.2. Global structure of the reduced system}

\noindent
The Hamiltonian flow on the manifold
$M^{\rm red}_{\pm e}$
is governed by the reduced Hamiltonian (\redham), which yields
the equation of motion,
$$
{{d^2 x}\over {dt^2}} + \mu e^{2x} = 0.
\eqn\eqm
$$
The flow is incomplete (singular)
if $\mu < 0$, as is intuitively clear from the rapidly decreasing
potential well in which the particle sinks.  For instance, 
the solution,
$x(t) = - \hbox{ln}(\cos t)$ which satisfies 
the initial condition $x(0) = 0$ and ${{dx}\over{dt}}(0) = 0$
and corresponds to $\mu = - 1$, blows up at $t = {{\pi} \over 2}$.
But since there is no singularity 
in the full reduced system, 
the incompleteness of the Hamiltonian flow that arises  
in the reduced system when $\mu < 0$ 
is just a manifestation of the fact that 
the particle may leave the submanifold
$\M^{\rm red}_{\pm e}\subset \M^{\rm red}$ at finite time.
More concretely, the trajectory
of the free particle on $G$ determined by  an initial value
$(\pm e^q,J_{\pm e})$ at $t=0$ as
$g(t) = \pm e^{t J_{\pm e}} e^q$
may leave the open submanifold $G_{\pm e}$, because 
the flow of the reduced system 
is obtained by projecting the original flow on
$\M^{\rm c}\subset \M$ to $\M^{\rm red}$.
Thus the singularity occurs when $g_{22}$ vanishes,
which corresponds to $q$ (or $x$) reaching infinity.
In this respect, one can say that
the embedding of $\M_{\pm e}^{\rm red}$ into $\M^{\rm red}$
provides
a regularization of the Liouville model with $\mu < 0$ where
the singular (blowing up) trajectories are glued together smoothly
at $x$ infinity.  If, on the other hand, $\mu > 0$, then the Hamiltonian
flow is complete, and the two Liouville models are completely 
disconnected from each other.  
Hereafter we confine ourselves to the case $\mu < 0$ where the two
Liouville models are connected, and put $\mu = -1$
for simplicity since $\mu$ can be freely rescaled by shifting $x$
in (\redham).  We also choose $\nu^+ = - \nu^- = 1$ for definiteness.
          
So far, we have analyzed the structure of the reduced system only
{\it locally}, using
the gauge fixing to identify the reduced system 
$(\M^{\rm red},\Omega^{\rm red}, H^{\rm red})$
as one containing a pair of Liouville models
glued together along lower dimensional submanifolds.
To furnish a tool to gain information on the {\it global} structure,
we wish to have a global cross section (gauge fixing) 
of the gauge orbits in $\M^{\rm c}$.  
Such a cross section is furnished by
the {\it Drinfeld-Sokolov gauge}, which is used 
in the context of generalized KdV equations [\DS] 
and $\cal W$-algebras 
(see, e.g., [\FORTW]). 
In our context we need to use it doubly for $J$ and $\tilde J$, and
for this reason we call our global gauge fixing 
`double DS gauge' here.

A double DS gauge is defined by requiring that
the two generators $J$ and $\tilde J$ of the symmetry 
be of the dual form,
$$
J(u) = \pmatrix{0&u_2\cr 1&0\cr},
\qquad
\tilde J(u) = \pmatrix{0& 1\cr u_2&0\cr}.
\eqn\dds
$$
Note that by definition
$J$ and $\tilde J$ are not quite independent (see
(\rcur)).  It is readily seen that the condition for $\tilde J$ 
in (\dds) is fulfilled if $J$ is of the form in (\dds)
and
$$
g(u)=\pmatrix{u_2u_4& -u_3\cr u_3& -u_4\cr}.
\eqn\ddsg
$$
These parameters $(u_2, u_3, u_4) \in \Real^3$ 
are subject to the condition,
$$ F(u) := \det g(u) =  u_3^2 - u_2 u_4^2 = 1,
\eqn\hypers
$$
which defines the hypersurface ${\cal S}$ 
in $\Real^3$ as a model of $\RM$.
It is straightforward to check that
$dF(u)\vert_{F(u)=1} \neq 0$,
which implies that (\hypers) gives
a regular hypersurface diffeomorphic to $\M^{\rm red}$.
Regarding the $u_i$ as gauge invariant functions on the
constrained manifold $\M^{\rm c}$, we find the Poisson brackets
$$
\{ u_4, u_3\}={u_4^2\over 2},
\qquad
\{ u_4,u_2\}=u_3,
\qquad
\{ u_3, u_2\}=u_4 u_2.
\eqn\pbu
$$
The dynamics of the reduced system is determined by the Hamiltonian
$$
H^{\rm red}(u)={1\over 2}\tr J^2(u)=u_2.
\eqn\hamredu
$$
The relationship to the local description of the reduced
system given earlier is established by noting that,
since $g_{22}(u) = - u_4$, the cells $M_e^{\rm red}$
and $M_{-e}^{\rm red}$ are represented by the domains $u_4 < 0$ and
$u_4 > 0$, respectively.

Having obtained a global picture of the reduced system,
we now see
how the singularity of the Liouville model 
gets regularized when the two models are glued together.
Consider the classical solution with a constant energy
$E$, whose trajectory is the curve obtained by 
intersecting the hypersurface ${\cal S}$ in 
(\hypers) with $u_2=E$.
The curve is a hyperbola or ellipse depending on the sign of 
the energy $E$ (see Fig. 2),
$$
u_3^2 - E u_4^2 =1.
\eqn\curve
$$
It is now clear that, for $E<0$, the motion of the particle
is periodic, which implies that the particle 
does pass the border
$u_4 = 0$ along the curve, 
and when it does so it gives rise to a singularity
in the solution when viewed as a local subsystem, although
the solution is perfectly regular when viewed as a global system.
It is worth noting that the hypersurface ${\cal S}$ is 
not simply connected. In fact, the loop given by,
say, the ellipse in (\curve) for some $E<0$ 
cannot be contracted to a point on the surface ${\cal S}$.


\noindent \secno=3 \meqno=1

\medskip\bigskip
\centerline
{\bf 3. Quantization of the reduced system}
\medskip
\noindent
We have learned that the classical reduced system
consists of two Liouville subsystems glued together 
between which the point particle oscillates if the energy 
is negative.
In this section we wish to define a quantum mechanics 
of the reduced system and examine if this global feature appears
at the quantum level as well.
In quantum mechanics observables are represented by 
self-adjoint
operators.
Here we consider the self-adjoint Hamiltonian of our reduced system, 
which is perhaps the most fundamental observable, as a crucial
ingredient to set up the quantum theory.
The standard procedure [\RS] for finding self-adjoint Hamiltonians
is to choose a suitable domain
where the quantum Hamiltonian, now takes some operator form, 
becomes a {\it symmetric} operator, and then find an extended 
domain where it becomes a {\it self-adjoint} operator.
(For a brief review of the general procedure, see Appendix A.)  
After this procedure, 
we shall find that there exists  
a probability flow between the two subsystems even though
all the wave functions vanish at the junction of the two 
Liouville subsystems, a fact that illustrates 
that the two subsystems are also connected at the quantum
level.

\medskip 
\noindent
{\bf 3.1. Hamiltonian as a symmetric operator}

\noindent
The basic problem for quantizing our reduced system
is that the reduced phase
space $\RM$ is not quite a cotangent bundle of some
configuration space, but a system of two cotangent bundles
nontrivially combined.
To take this feature into account, we wish to formulate the
quantum theory dealing with the two subsystems
simultaneously, in such a way that the classical
connectedness of the two subsystems will also be realized
quantum mechanically.

In section 2 we have seen that there exists a global description
of the reduced system using the hypersurface ${\cal S}$ as a model
for the reduced phase space.  Among the three parameters used we  
choose the variable 
$$
Q := g_{22}(u) = - u_4
\eqn\q
$$
and regard it as `coordinate' of the particle.  The variable 
is chosen on the ground that the component $g_{22}$ is gauge invariant
under the symmetry action (\naction) and hence from (\decomp)
we have the direct 
relation $Q = \pm e^{-x}$ with the local Liouville
coordinate $x$.  The variable $Q$ is convenient for our purpose since
$Q > 0$ corresponds to the subsystem 
$(\RM_{e}, \Omega^{\rm red}_{e}, H_{e}^{\rm red})$ and $Q < 0$
to $(\RM_{-e}, \Omega^{\rm red}_{-e}, H_{-e}^{\rm red})$, respectively.
The canonical momentum conjugate to $Q$ is then given by 
$P := - 2 u_3 / u_4^2$ which satisfies
$$
\{ Q\, , \, P \} = 1,
\eqn\pc
$$
under the reduced Poisson brackets (\pbu).
In terms of these variables the classical reduced Hamiltonian 
(\hamredu) reads
$$
H= \frac{1}{4}Q^2 P^2 - {1\over{Q^2}}.
\eqn\hc
$$
Upon the identification $P = 2 e^x \pi$ 
the Hamiltonian (\hc) agrees with the local expression (\redham).
Now the trouble is that the Hamiltonian is ill-defined at
$Q = 0$, i.e., at the junction between the
two subsystems.  We exclude this point from the
domain where $Q$ is defined:
$Q \in \Real^* := \Real \backslash \{0\} = \Real^+ \cup \Real^- $.
As we shall see soon, despite this exclusion and the apparent
trivialization of the reduced system into two decoupled subsystems, 
it is possible to construct 
a quantum theory such that the two Liouville 
subsystems are connected nonetheless. 

In quantization we elevate these canonical 
variables to linear operators on a Hilbert space with
the Poisson bracket (\pc) replaced by the commutation relation,
$$
 [\hat Q\, , \, \hat P] = i.
\eqn\pq
$$
Working with the coordinate representation,  
we define the Hilbert space by the space 
of square integrable functions,
$$
{\cal H} := \bigl\{ \phi \, \big\vert \, \| \phi \| 
< \infty \bigr\},
\eqn\hs
$$
where $\| \phi \| = \sqrt{\langle \phi,\phi \rangle}$ is the norm 
of the state $\phi$.  We furnish the innerproduct by 
$$
\langle \phi,\psi \rangle 
= \langle \phi,\psi \rangle_- + \langle \phi,\psi \rangle_+,
\eqn\no
$$
where
$$
\langle \phi,\psi \rangle_-
        :=\int_{-\infty}^{-0}\frac{dQ}{|Q|}\, 
\phi(Q)^{\ast}\psi(Q), 
\qquad
\langle \phi,\psi \rangle_+ 
        :=\int_{+0}^{+\infty} \frac{dQ}{|Q|}\, 
\phi(Q)^{\ast}\psi(Q).
\eqn\innerpdt
$$
The measure $dQ/|Q|$ used in the innerproduct (\innerpdt)
can be derived, for instance, by the path-integral reduction, 
where the original measure $\prod_t\Omega^3(t)$
for the free particle
system on $SL(2, \Real)$ reduces to the form $\prod_t dQ/|Q|(t)$ 
after we integrate out the momentum variables $J$ 
taking into account the constraints and the gauge fixing conditions
[\ORT].  A more direct way to see this is 
to consider the phase
space path-integral for the reduced system with the Hamiltonian   
(\hc) and then integrate on $P$ to get the configuration space
path-integral, which results precisely in the measure $dQ/|Q|$.
Note that the measure $dQ/|Q|$ is just 
the standard Toda measure
$dx$ in terms of the local coordinate $x$, as expected.  
Note also that, because of the measure, 
all the wave functions $\phi(Q) \in {\cal H}$ must
vanish at the junction of the two subsystems: $\phi(Q) \rightarrow 0$
as $Q \rightarrow \pm 0$.

In order to find a self-adjoint Hamiltonian operator, 
let us consider the
differential operator $\hat H$ of the form, 
$$
\hat {H} :=  -\frac{1}{4}Q \frac{d}{dQ} 
Q \frac{d}{dQ} - \frac{1}{Q^2},
\eqn\hq
$$
which is a naive choice for the 
operator which corresponds to the 
classical Hamiltonian
(\hc).  We can find a domain where the Hamiltonian operator 
is symmetric,
$$
D(\hat {H} ) := \bigl\{ \psi\, \big\vert \, 
\psi \in {\cal H}, \, \hat {H}\psi \in {\cal H}, \,
\lim_{Q \rightarrow \pm 0, \pm \infty}Q\psi(Q) = 
Q \frac{d}{dQ}\psi(Q)=0 \bigr\}.
\eqn\hsd
$$
In fact, it can be readily confirmed
that on $D(\hat {H})$ we have
$$
\< \hat {H} \psi_1,\psi_2 \> = 
\< \psi_1 ,\hat {H} \psi_2 \> 
\qquad\quad \forall\, \psi_1,
\psi_2 \in D(\hat {H} ).
\eqn\hs
$$
It is also easy to see that the domain 
of the adjoint operator $\hat {H}^{*}$, which 
as a differential operator takes the same form 
as $\hat H$,  
is just $D(\hat {H}^*) := \bigl\{ \psi\, \big\vert \,
\psi \in {\cal H}, \, \hat {H}\psi \in {\cal H}
\bigr\}$.  This shows that the symmetric operator
$\hat H$ is not self-adjoint; $D(\hat H^*) \supset D(\hat H)$.

Before discussing the self-adjoint extensions of the symmetric operator,
let us consider the eigenvalue problem of the differential operator 
$\hat {H}$,
$$
\hat {H} \phi = E \phi.
\eqn\hei
$$
Using the variable
$$
z=\frac{2}{Q},
\eqn\tr
$$
we find that eq.(\hei) becomes 
$$
\left[ z^2 \frac{d^2}{dz^2}
+ z \frac{d}{dz} + (z^2-k^2)\right] \phi = 0,
\eqn\ee
$$
where
$$
k^2=-4E.
\eqn\ie
$$
The linearly independent eigenstates of eq.(\hei) are given by 
the Bessel functions $J_k(z)$ and $Y_k(z)$ with indices 
$k \in \Complex $.  However, only those Bessel functions of the type
$J_k(z)$ with ${\rm Re}\,k>0$
have finite norms and belong to the Hilbert space 
${\cal H}$.  In particular, for $E < 0$ there exists a unique
$k$ for which the corresponding eigenstate belongs to 
${\cal H}$ while there is 
no such eigenstate belonging to ${\cal H}$ for $E \geq 0$, 
and for this reason we henceforth
restrict ourselves to the negative energy states.
Note however that {\it none} of 
these negative energy eigenstates  
belongs to the domain $D(\hat H)$ where the Hamiltonian is symmetric,
as can be seen from their asymptotic forms (see Appendix B),
$$
\eqalign
{
QJ_{k}(\frac{2}{Q})&~~^{~~\sim}_{Q \rightarrow +0} 
~~\frac{1}{\sqrt{\pi}}\,Q^{\frac{3}{2}}
\cos\bigl(\frac{2}{Q}-\frac{\pi}{2}(k+\frac{1}{2})\bigr), \cr
Q\frac{d}{dQ}J_{k}(\frac{2}{Q})&~~^{~~\sim}_{Q \rightarrow +0}~~
-\frac{1}{\sqrt{\pi}}\, Q^{-\frac{1}{2}}
\sin\bigl(\frac{2}{Q}-\frac{\pi}{2}(k+\frac{1}{2})\bigr).
}
\eqn\as
$$
Clearly, as $Q$ tends to 
zero the first term $QJ_{k}(2/Q)$ goes to zero while
the second term $Q\frac{d}{dQ}J_{k}(2/Q)$ blows up to infinity, 
showing that $J_k(z) \not \in D(\hat {H})$ for Re$\,k > 0$.

\def\H{\hat H}
\medskip
\noindent
{\bf 3.2. Self-adjoint extensions}

\noindent
In order to find an extended domain where
the symmetric operator $\hat H$ becomes a self-adjoint operator 
$\hat H_*$ for which 
$D(\hat H_*) = D(\hat H_*^*)$,
we first examine, according to the general theory (see Appendix A),
the deficiency indices $(d_+, d_-)$ of the symmetric operator $\hat H$.
The index $d_+$ ($d_-$) is given by the 
dimension of the eigenspace of the adjoint operator $\hat H^*$
with eigenvalue $i$ ($-i$),
$$
d_{+} :=  \hbox{dim}\, \hbox{Ker} (\H^{\ast}-i), \qquad
d_{-} :=  \hbox{dim}\, \hbox{Ker} (\H^{\ast}+i).
\eqn\di
$$
We recall that  
the eigenfunctions of the differential operator $\hat H$ in (\hq) 
with eigenvalue $i$ ($-i$)
are given by the Bessel functions 
$J_{\pm k_0}$ ($J_{\pm k_0^{\ast}}$) 
with index $k_0 := 2 \exp (-\frac{\pi}{4}\!i)$. 
But since 
for these eigenfunctions to be in the Hilbert space ${\cal H}$
the real part of the indices $\pm k_0$ ($\pm k^*_0$) must be positive,
the only eigenstate allowed is $J_{k_0}$ ($J_{k_0^{\ast}}$), which
actually belongs to the domain of the adjoint operator $D(\hat H^{\ast})$. 
We therefore observe that $(d_+, d_-) = (1, 1)$, a result ensuring that
$\hat H$ admits self-adjoint extensions.

The general theory of self-adjoint extension then asserts that
the extended domain where the 
symmetric operator becomes self-adjoint
consists of three spaces, i.e., 
the domain of the symmetric operator $D(\hat H)$, 
the eigenspace
of the adjoint operator $\H^*$ with eigenvalue $i$, and the eigenspace of
$\H^*$ with eigenvalue $-i$, with the latter two spaces being
related unitarily.  Concretely, it is given by
$$
D(\hat H_*)=\bigl\{ \psi \, \big\vert \, \,
\psi = \phi + \alpha \xi_{k_0}, \,\,
\phi \in D(\hat H),\, \alpha \in \Complex
\bigr\},
\eqn\ds
$$ 
where 
$$
\xi_{k_0} := \frac{J_{k_0}}{\| J_{k_0} \|}+ e^{i\theta}
\frac{J_{k^*_0}}{\| J_{k^*_0} \|}.
\eqn\func
$$
The point to be noted is that the domain is parametrized by 
the angle,
$$
\theta \in [~0,2\pi).
\eqn\co
$$
The angle parameter $\theta$ 
therefore characterizes the quantum theory we 
construct, whose existence is just a reflection of the 
ambiguity in quantizing a classical system.

Having achieved the self-adjoint extensions of the Hamiltonian, 
we now 
investigate which eigenstates $J_{k}$
of the adjoint operator
belong to $D(\hat H_*)$ in (\ds) for a specific $\theta$. 
A necessary and sufficient condition for  
$J_{k} \in D(\hat H_*)$ is [\N]
$$
\<\hat H^* J_{k}, \psi \>= \< J_{k},\hat H_* \psi \> 
\qquad\quad \forall \, \psi \in D(\hat H_*).
\eqn\si
$$
which takes the form
$$
Q\,W(J_{k}^{\ast},\psi)\Big\vert^{-0}_{-\infty} -
Q\,W(J_{k}^{\ast},\psi)\Big\vert^{+\infty}_{+0}=0,
\eqn\bc
$$
where $W(\phi_1, \phi_2) = \frac{d\phi_1}{dQ}\phi_2
                  - \phi_1\frac{d\phi_2}{dQ}$
is the Wronskian.
An equivalent condition is obtained if we replace $\psi$ with 
$\xi_{k_0}$ in (\bc), and by using the asymptotic forms (\as)
we arrive at the relation
$$
e^{2\pi i (k-{\rm Re}\,k_0)} 
= {{\cos{\theta\over2}\cosh{(\pi{\rm Im}\,k_0)}
   - i\sin{\theta\over2}\sinh{(\pi{\rm Im}\,k_0)}}\over
   {{\cos{\theta\over2}\cosh{(\pi{\rm Im}\,k_0)}
    + i\sin{\theta\over2}\sinh{(\pi{\rm Im}\,k_0)}}}}.
\eqn\wrr
$$
Solving this relation in favour of $k$, we get
$$
k = k(n,\theta) := {\rm Re}\, k_0 - {1\over\pi} \tan^{-1}
\left[\tan{\frac{\theta}{2}}\tanh{(\pi{\rm Im}\,k_0)}\right] + n,
\eqn\ei
$$
where $n$ are integers for which $k > 0$.
We therefore see that the eigenstates 
allowed by the self-adjoint Hamiltonian labelled by $\theta$ 
are characterized by the indices 
$k(n,\theta)$ whose energy eigenvalues are  
discrete, and that from (\ie)
the intervals of adjacent discrete energy levels 
depend on the angle parameter.
Note that as we vary the parameter  
from $\theta = 0$ 
the spectrum changes accordingly and returns to the original
spectrum only when $\theta$ approaches to $2\pi$.
Since any two Bessel functions whose indices differ by 
an integer are orthogonal to each other with respect to
the innerproduct on $\Real^*$,
so are any of the two eigenstates in the domain 
$D(\hat H_*)$, as required.

\medskip
\noindent
{\bf 3.3. Probability flow between the two Liouville subsystems}

\noindent
We have seen in section 2 that the reduced 
classical system admits solutions $Q(t)$ oscillating 
between the two subsystems given by $Q > 0$ and $Q < 0$.
We now analyze how this global aspect manifests itself
in the quantum theory.  More specifically, we are interested
in the question if there exists a probability flow between 
the two subsystems.   We shall find that the answer 
is positive, signaling the fact that the reduced 
system is a connected system also quantum mechanically.   

{}For this, consider the state $\psi(t)$ given by a linear 
combination of two energy eigenstates 
$$
\psi(t)= c_1J_{k_1}({2\over{Q}}) 
e^{-iE_{k_1}t}+c_2J_{k_2}({2\over{Q}})e^{-iE_{k_2}t},
\eqn\wf
$$
where 
$J_{k_1}, J_{k_2}\in D(\hat H_*)$ for some fixed $\theta$
and $E_{k_i} = - k_i^2/4$
for $i = 1$, 2.  We take the two indices $k_1$ and $k_2$ 
which are different by some odd integer,
$$	
k_1-k_2=2n+1, \qquad n \in \Zed,
\eqn\oddint
$$
and choose the constants $c_1$, $c_2 \in \Complex$ so that
the state be normalized $\| \psi \| = 1$ at $t = 0$.
Then the orthogonality condition $\langle J_{k_1},J_{k_2}\rangle = 0$
implies that the norm, i.e., the total
probability on the full line $\Real^*$, remains constant, 
$$
\frac{d}{dt} \langle \psi,\psi \rangle 
= \frac{d}{dt} \langle \psi,\psi \rangle_+
    + \frac{d}{dt} \langle \psi,\psi \rangle_- = 0. 
\eqn\wftd
$$
However, the probability 
on a half line, say $\Real^+$, does not
remain constant.  Indeed,
a similar computation for the half line reveals that
$$
\frac{d}{dt}\langle\psi,\psi\rangle_+ 
 = i c_1^{\ast}c_2 \langle J_{k_1}, J_{k_2}\rangle_+
 (E_{k_1}-E_{k_2})e^{-i(E_{k_2}-E_{k_1})t} + \hbox{c.c.}\, \neq 0
\eqn\wftdd
$$
because 
the two Bessel functions with (\oddint) 
are not orthogonal to each other
$\langle J_{k_1},J_{k_2}\rangle_+ \neq 0$
on the half line $\Real^+$. 
This shows that there exists a probability flow between 
the two subsystems where $Q \in \Real^{+}$ and $Q \in \Real^{-}$,
even though the wavefunctions vanish at the junction.
The reason why such a flow can exist is that 
the Hamiltonian operator 
is {\it not} self-adjoint with respect to the innerproduct 
on the half line $\Real^+$ (or $\Real^-$), although it becomes
so if we cut the domain `in half' so that it consists only of 
those eigenstates with indices given by even (or odd) integers
$n$ in (\ei).

\bigskip
\medskip
\centerline
{\bf 4. Discussion}
\medskip
\noindent
We have seen in this paper that the quantum mechanical Liouville
model with attractive potential 
obtained by the Hamiltonian reduction (which regularizes
the Liouville model classically) 
can be solved completely.  Although the energy
spectrum is unbounded from below, the fact that only discrete 
levels are allowed suggests that the system is `quasi-stable'
at the quantum level.  The connectedness of
the two subsystems can be observed by a probability flow, 
which we have shown to exist. 
The quantum theory is labelled
by the angle parameter $\theta$ which arises in constructing
self-adjoint Hamiltonian operators.  In this respect, 
it is worth noting that
the reduced phase space is topologically $\Real^2\backslash\{0\}$,
that is, the two dimensional Euclidean 
plane with a hole, which is 
homeomorphic to $T^*S^1 = S^1 \times \Real$.  The appearance
of the angle parameter may perhaps be understood as a
common phenomenon observed in quantizing on a configuration space with
a hole, as in the case of the quantum
theory on $S^1$ or of the Yang-Mills theory, whose quantization yields
the $\theta$-vacua by an analogous mechanism.
 
Our procedure of quantization is similar to that of F\"ul\"op
[\F] but differs in some important points.  First, 
unlike our measure $dQ/|Q|$ for the innerproduct (\innerpdt), 
the measure used for the innerproduct in [\F] is $|Q| dQ$,
which is obtained from the Haar measure of the group $SL(2,\Real)$
by eliminating the degrees of freedom that correspond to
the symmetry.  However, 
in Hamiltonian reduction one has to use the phase space volume 
element on $T^*SL(2,\Real)$ 
to derive the correct reduced measure [\ORT], which is 
the measure $dQ/|Q|$ we used. 
This causes a certain alteration in the energy spectrum.  
Second, and more importantly, 
the self-adjoint extensions of the Hamiltonian operator 
are achieved in [\F] by considering a domain such that 
the eigenstates on the full line $\Real^*$ 
are formed out of linear combinations of two
eigenstates, each defined on the half lines
$\Real^+$ and $\Real^-$, respectively.
Since one can take
distinct angle parameters to specify the self-adjoint extensions
on the two half lines, one needs two angle
parameters in general to specify the self-adjoint extensions on the full
line $\Real^*$ (plus an extra parameter to render the eigenstates
mutually orthogonal).  
This we find an unnecessary complication,
given that the self-adjoint extensions can be achieved
on the full line without referring to those on the half lines.
Our simpler quantization yields
the energy spectrum given by a single class of discrete levels 
with indices (\ei) specified by the angle $\theta$, a result
which we feel is natural to associate with the classical 
periodic motions on the smooth phase space.
In contrast, 
the spectrum in [\F] 
consists of two classes of discrete levels similar to (\ei)
but with even integers $n \in 2 \Zed$. 

{}Finally, we wish to stress that the quantization
discussed in this paper is not the unique one available 
to the Liouville system.
Indeed, from the way the classical system is defined, it is perhaps
more natural to quantize first the system
of a free particle on $G$ and then carry out quantum
Hamiltonian reduction.  This will provide
a way to confirm what we have learned in the quantum mechanical 
Liouville model given in this paper.

\bigskip
\noindent{\bf Acknowledgement}

\noindent
We wish to thank S. Tsujimaru for useful discussions in the early
stages of the work.  We also thank L. Feh\'er for valuable comments.

\bigskip


\noindent \secno=0 \appno=1 \meqno=1

\def\a{\hat A}
\def\b{\hat B}
\def\kp{K_+(\a^*)}
\def\km{K_-(\a^*)}
\noindent
{\bf Appendix A. Symmetric operator and self-adjoint extension}
\medskip
\noindent
Here we briefly summarize the general theory of
self-adjoint extensions of symmetric operators [\RS].
Let $\hat A$ be a linear operator on a dense domain $D(\a)$ in
a Hilbert space $\cal{H}$.  Consider $\psi \in {\cal H}$
for which 
$$
\<\psi, \a\, \phi\>=\<\psi', \phi\> \qquad \forall\, \phi \in D(\a).
\eqn\inn
$$
is satisfied for some $\psi' \in {\cal H}$.  Define the domain
$D(\hat A^*)$ by the set consisting of those $\psi$.  It is then 
easily confirmed that $\psi \mapsto \psi'$ gives a linear map.
The {\it adjoint} operator 
$\a^{\ast}$ is defined by  
this linear map $\a^{\ast}$: $\psi' = \a^{\ast}\psi$.
If the operator $\a$ fulfills the condition
$$
\qquad \,D(\a^{\ast}) \supset D(\a),\qquad \a^{\ast} = 
\a \quad \hbox{on} \quad D(\a),
\eqn\dad
$$
then $\a$ is a {\it symmetric} operator.
A symmetric operator is called a {\it self-adjoint} operator
when the two domains coincide, $D(\a^{\ast}) = D(\a)$.
If $\a$, $\b$ are two operators such that
$$
D(\b) \supset D(\a),
\qquad \b = \a \quad \hbox{on} \quad D(\a),
\eqn\res
$$
then the operator $\b$ is an {\it extension} 
of the operator $\a$.  A symmetric operator $\a$ can be extended
to be a self-adjoint operator $\a_*$ 
if there exists a domain $D(\a_*)$ such that
$$
D(\a^*) \supset D(\a_*^*) = D(\a_*) \supset D(\a).
\eqn\domains
$$ 

When a certain condition is fulfilled (which we discuss shortly), 
such {\it self-adjoint extensions} of a symmetric operator $\a$
are possible in the following way.
We begin by decomposing 
the domain $D(\a^*)$ of the adjoint operator $\a^*$ as
$$
D(\a^*) =  
D(\a) + K_-(\a^*) + K_+(\a^*), \qquad\hbox{where}\quad
K_\pm(\a^*) := \hbox{Ker}\,(\a^{\ast}\pm i).
\eqn\deco
$$
To see that this decomposition is possible,
we first decompose
any state $\psi \in D(\a^{\ast})$
as $\psi = \zeta + \alpha$ where
$\zeta \in D(\a)$ and $\alpha \not\in D(\a)$.  Applying
$(\a^{\ast}-i)$ to $\psi$ we find
$$
(\a^{\ast}-i)\psi = (\a-i)\zeta +(\a^{\ast}-i)\alpha.
\eqn\dd
$$
Note that 
on account of the property $\|(\a \pm i)\phi\|^2 = \|\a\phi\|^2
+ \|\phi\|^2$ for $\forall\, \phi \in D(\a)$
the spaces $(\a \pm i)D(\a)$ are closed subspaces in ${\cal H}$,
and that they are
orthogonal to $K_\mp(\a^*)$, respectively.
Thus we can write
$(\a^{\ast}-i)\alpha = (\a - i)\beta - 2i\xi$,
where $\beta \in D(\a)$ and 
$\xi \in \kp$.  But since 
$ -2i\xi = (\a^* - i)\xi$, we have
$$
(\a^{\ast}-i)(\psi -\phi-\xi)=0,
\eqn\dofadd
$$
where $\phi := \zeta + \beta \in D(\a)$.
It then follows that $\eta := \psi -\phi -\xi \in 
\km$, that is, any state $\psi \in D(\a^*)$
can be decomposed as
$$
\psi = \phi + \xi + \eta, \qquad 
\phi \in D(\a),\quad
\xi \in \km,\quad
\eta \in \kp,
\eqn\condd
$$
which proves our claim (\deco).

It can be shown that 
a necessary and sufficient condition for satisfying 
the relation
$$
\<\psi, \a^{\ast}\psi\>=\<\a^{\ast}\psi, \psi\>  
\eqn\cond
$$
for $\psi \in D(\a^*)$ in the form (\condd)
is $\|\xi\|=\|\eta\|$, i.e., the latter two components must be 
related by a unitary transformation $\eta = \hat U \xi$.
Such a unitary transformation exists if and only if
the two {\it deficiency indices}
$(d_{+},d_{-})$ defined by
$$
d_+ = \hbox{dim}\, \km,\qquad 
d_- = \hbox{dim}\, \kp,
\eqn\di
$$
are equal $d_+ = d_-$.  Hence, if this is the case,  
the symmetric operator $\a$ can be 
extended to a self-adjoint operator $\a_*$ with the domain
$$
D(\a_*)=\{\psi\, |\, \psi = \phi + \xi + \hat U \xi, \quad
\phi \in D(\a), \quad
\xi \in \km, \quad
\hat U \xi \in \kp\}.
\eqn\adomain
$$


\noindent \secno=0 \appno=2 \meqno=1

\medskip
\noindent
{\bf Appendix B. Orthogonality and the 
asymptotic forms of Bessel functions}
\medskip
\noindent
Bessel functions $J_k(z)$ are defined by the series,
$$
J_{k}(z)=\sum_{m=0}^{\infty}
\frac{(-1)^m}{m!\,\Gamma (k +m+1)}\left(\frac{z}{2}\right)^{k +2m},
\qquad \hbox{arg}\,|z|< \pi, \quad k \in \Complex.
\eqn\defbe
$$
The other type of Bessel functions are 
$Y_k(z) := (\cos(k\pi) J_k(z) - J_{-k}(z))/\sin(k\pi)$.
The Bessel functions (\defbe) have the following asymptotic forms
for $z$ approaching infinity and zero:
$$
\eqalign{
J_{k}(z)&~~~^{~~\sim}_{z \rightarrow +\infty}~~~ 
\sqrt{\frac{2}{\pi z}} \cos [z-\frac{\pi}{2}(k +1)], \cr 
J_{k}(z)&~~~^{~~\sim}_{z \rightarrow +0} ~~~
\frac{1}{\Gamma(k+1)}\left(\frac{z}{2}\right)^{k}.
} 
\eqn\asz
$$
We define the Bessel functions on $\Real^{-}$
by analytically continuing them 
from the upper half plane $\hbox{Im}z > 0$, that is, we give the 
values on $\Real^{-}$ by rotating the functions on $\Real^+$
by $\pi$,
$$
J_{k}(z) = J_{k}(e^{\pi i}(-z)) = e^{k \pi i}J_{k}(-z),
\qquad \hbox{arg}\,z=\pi.
\eqn\anacon
$$
Note that $J^*_k(z) = J_{k^*}(z)$ for $z \in \Real^+$ whereas
$J^*_k(z) = e^{-2i k^*\pi} J_{k^*}(z)$ for $z \in \Real^-$.

Due to the formula (\anacon), integrals of the Bessel functions
on the full line $\Real\backslash \{0\}$ read
$$
\langle J_{k},J_{l} \rangle =
\bigl(1+e^{-i(k^{\ast}-l)\pi}\bigr) 
\langle J_{k},J_{l} \rangle_+.
\eqn\ortho
$$
The integrals $\<J_{k},J_{l}\>_+$
on the half line $\Real^+$ can be 
evaluated by means of Lommel's integral [\W],
$$
\int_{a}^{b}\frac{dz}{z}J_{k}(z)J_{l}(z)=
\frac{1}{k^2-l^2}
\Bigl[z\bigl(J_{k}(z)\frac{d}{dz}J_{l}(z)-
J_{l}(z)\frac{d}{dz}J_{k}(z)\bigr)\Bigr]^{b}_{a}.
\eqn\lom
$$
If $\hbox{Re}\, (k +l) >0$ then the above integral converges, and
from $Q=2 / z$ we obtain
$$
\<J_{k},J_{l}\>_+ =
\lim_{\scriptstyle a\rightarrow +0\atop \scriptstyle
b\rightarrow +\infty}
\int_{a}^{b}\frac{dQ}{\vert Q \vert}J_{k}^{\ast}(\frac{2}{Q})J_{l}
(\frac{2}{Q}) 
=\frac{2\sqrt 2}{\pi(k^{\ast 2}-l^2)} \sin
[\frac{\pi}{2}(k^{\ast}-l)],
\eqn\orthoo
$$
where we used the asymptotic forms (\asz).
If $k$, $l \in \Real$ and $k - l = n \in \Zed$ then
the integral on the full line vanishes, $\<J_{k},J_{l}\> = 0$.
This occurs for two different reasons depending on whether
$n$ is even or odd; 
for even $n$, the integral on the half line vanishes
$\<J_{k},J_{l}\>_+ = 0$, whereas for odd $n$ 
the integral is not zero but 
the factor $(1+e^{-i(k^{\ast}-l)})$ vanishes.

  \vfill\eject\immediate\closeout\reffile
  \centerline{{\bf References}}\bigskip
\baselineskip 15pt
  \frenchspacing%
  \input refs.tmp\vfill\eject\nonfrenchspacing


\vfill\eject

\centerline{\bf Figure Captions}

\vskip 1cm

\noindent
{\bf Figure 1.} A schematic picture of
the potential in the regularized Liouville system.
\vskip 7mm
\noindent
{\bf Figure 2.} Two typical trajectories 
on the reduced phase space.  The ellipse ($E < 0$)
corresponds to a periodic motion whereas the hyperbola 
($E > 0$) corresponds 
to a motion passing once for all from one cell to another.

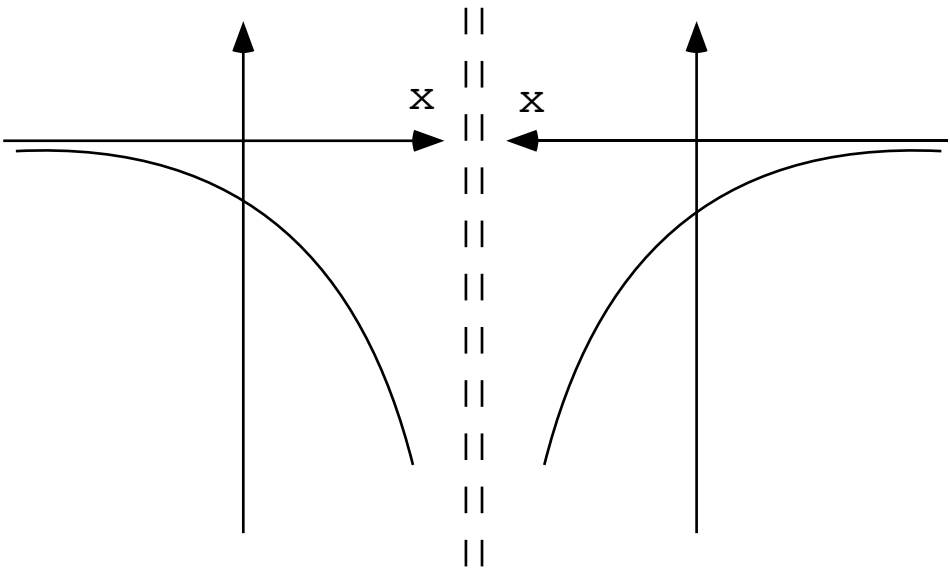
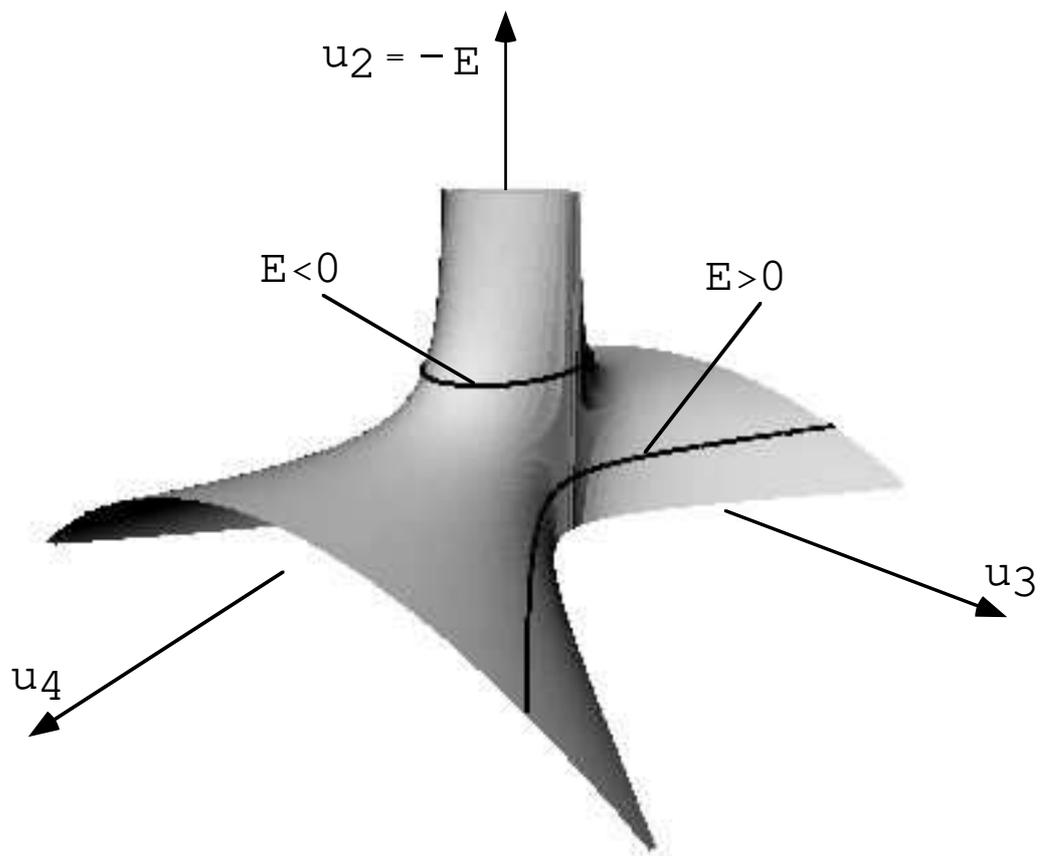
\bye